\documentclass[journal,draftclsnofoot,onecolumn,12pt]{IEEEtran}
\usepackage{amsmath}
\usepackage{balance}
\usepackage{cite}
\usepackage{hyperref}
\usepackage{amsfonts}
\usepackage{amsmath}
\usepackage{amssymb}
\usepackage[usenames]{color}
\usepackage{balance}
\usepackage{amsmath}
\usepackage{amssymb}
\usepackage{multirow}
\usepackage{cite}
\usepackage{floatrow}
\newfloatcommand{capbtabbox}{table}[][\FBwidth]

\usepackage{subfig}
\usepackage{graphicx}

\begin{document}

\title{Analysis of a Frequency-Hopping \\Millimeter-Wave Cellular Uplink}
\author{Don Torrieri,~\IEEEmembership{Senior~Member,~IEEE,} Salvatore Talarico,~\IEEEmembership{Member,~IEEE,}\\and Matthew C. Valenti,~\IEEEmembership{Senior~Member,~IEEE.} \thanks{Portions
of this paper were presented at the 2015 IEEE Military Communications Conference \cite{tsv}. D. ~Torrieri is an electrical engineer and mathematician
(email: dtorrieri@verizon.net). S.~Talarico was with West Virginia University, Morgantown, WV, U.S.A, and he is now with Huawei Technologies, Santa Clara, CA, U.S.A (email: salvatore.talarico81@gmail.com). M. C. Valenti is with West Virginia University, Morgantown, WV, U.S.A (email: valenti@ieee.org).} \vspace{-1cm} }
\maketitle

\begin{abstract}
\normalsize
Fifth-generation (5G) cellular networks are expected to exhibit at least three primary physical-layer differences relative to fourth-generation ones: millimeter-wave propagation, massive antenna arrays, and densification of base stations. As in fourth-generation systems, such as LTE, 5G systems are likely to continue to use single-carrier frequency-division multiple-access (SC-FDMA) on the uplink due to its advantageous peak-to-average power ratio. Moreover, 5G systems are likely to use frequency hopping on the uplink to help randomize interference and provide diversity against frequency-selective fading. In this paper, the implications of {these and other physical-layer features on uplink performance are assessed using a novel millimeter-wave propagation model featuring distance-dependent parameters that characterize the path-loss, shadowing, and fading.  The analysis proceeds by first fixing the location of the mobile devices and finding the performance conditioned on the topology.  The spatially averaged performance is then found by averaging with respect to the location of the mobile devices.  The analysis allows for the use of actual base-station topologies and the propagation model can leverage empirical millimeter-wave measurements.} The benefits of base-station densification, highly directional sectorization, frequency hopping, a large available bandwidth, and a high code rate are illustrated. The minor importance of fractional power control is shown.
\end{abstract}


\section{Introduction}

It is considered likely that 5G cellular networks will operate in millimeter-wave bands \cite{andr,wang}.
However, the basic structure of the fourth-generation (4G) single-carrier frequency-division
multiple-access (SC-FDMA) uplink systems will likely be maintained due to the critical importance of transmitter power efficiency, which is enabled by the low peak-to-average power ratio of SC-FDMA and similar
single-channel modulations \cite{deb,ali}. It is also likely that 5G systems
will exploit frequency hopping to compensate for frequency-selective fading {and to randomize out-of-cell interference},
as is done in 4G systems \cite{ku}.
{Other likely features of 5G networks include the use of multi-tier heterogenous networks (HetNets) and the decoupling of the uplink and downlink \cite{boc}.  On the downlink, biasing is used in a HetNet to expand the range of small cells, thereby balancing cell loads.  When the uplink and downlink are coupled, such biasing can cause a mobile device to transmit to a distant base station, which is not optimal from the uplink perspective.  However, if the uplink and downlink are decoupled, then the mobile device is free to transmit to a closer base station.  Because of blockages and severe shadowing, the base station that receives the strongest signal from a mobile device is not necessarily the one that is closest to it.  Thus, in a millimeter-wave network with a decoupled uplink, it is advantageous for the mobile device to establish an uplink with the base station that receives its signal with the greatest strength.}

This paper provides a performance analysis of a frequency-hopping millimeter-wave uplink that
can serve as the uplink of a fifth-generation (5G) cellular network in its
primary mode of operation. The work contained in this paper further elaborates and improves on the analysis of
\cite{tsv}.  {To drive the analysis, a novel millimeter-wave propagation model is proposed with distance-dependent parameters that characterize the path-loss, shadowing, and fading.  The} distance dependence of these models accounts for the fact that mobiles close to the BS have a line-of-sight (LOS) path, but the more distant mobiles do not.   One of the main goals is to examine the impact of base station (BS) or cell
densification, which is the increase in the number of BSs or cells relative to
the number of mobiles served. A major benefit of densification is to bring
each mobile closer to its serving BS. Another benefit of densification is that
it allows each mobile in a cell access to a larger portion of the available
spectrum without causing intrasector interference.

The analysis proceeds by deriving the \emph{conditional} outage probability of the uplink, where the
conditioning is with respect to an arbitrary network topology. In the
numerical examples, we use an actual BS deployment.
Each simulation trial generates a network realization in which the mobile
placements are drawn from the uniform clustering distribution \cite{torval1},
while the locations of the BSs remain fixed. For each realization of the
network, the outage probability is computed by using closed-form equations for
a reference link. By averaging over many mobile station placements, the
average outage probability and other statistical performance measures are
computed for the given BS deployment. The benefit of this approach relative to
other approaches based on stochastic geometry is that it captures the fact
that mobiles do not maintain fixed positions, but BSs do.
The analysis in this paper applies the methodology and model of \cite{torval1}%
, which we call \emph{deterministic} geometry to contrast it with stochastic
geometry \cite{hae,dire}.
The two approaches are connected in \cite{val}, which proves that deterministic
geometry yields the same spatial averages as stochastic geometry.
Deterministic geometry can accommodate arbitrary finite topologies with distance-dependent
path loss, shadowing, and fading models without making the many {typical} approximations of stochastic geometry, such as assuming that the fading is the same for all links,
{or that the topology is drawn from a Poisson point process (PPP) or determinantal point process (DPP)\cite{li}.}
Moreover, deterministic geometry is able to seamlessly capture the effects
of thermal noise, which will be shown to be more detrimental than the
interference, and is included in the model without any
approximation.

The remainder of the paper is as follows. Section \ref{sec:Network_model}
presents the network model, which includes descriptions of the network
topology, the propagation model, sectorization and beamforming, frequency
hopping, intrasector and intersector interference, spectral assignments, and
power control. Section \ref{sec:Outage_Probability} presents the equations
governing the calculation of the conditional outage probability. Some
numerical evaluations for a typical 5G cellular network are given in Section
\ref{sec:Results}. Conclusions are drawn in Section \ref{sec:Conclusions}.

\section{Network Model}

\label{sec:Network_model}

\subsection{Network Topology}

\begin{figure}[tb]
\centering
\includegraphics[width=10.5 cm]{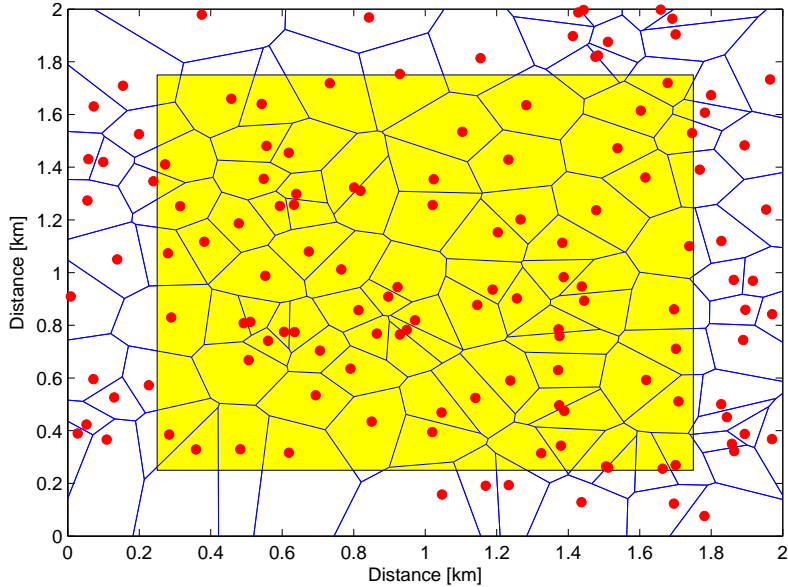}
\caption{Actual BS locations from a current cellular deployment. BSs are represented by large circles, and Voronoi cell boundaries are represented by thick lines. To minimize edge effects, the reference mobile is assumed to be located in the yellow zone in the center of the diagram.}%
\label{fig:Figure_1}
\end{figure}

In the network model, $C$ BSs and $M$ mobiles are confined to a finite area.
As an example, Fig. \ref{fig:Figure_1} depicts an actual deployment of
$C=132$ BSs extracted from the OFCOM database for Vodafone in London{\footnote{{http://www.sitefinder.ofcom.org.uk}}}.
BSs are represented by large circles, and Voronoi cell boundaries are
represented by thick lines. The network occupies 4 km$^{2}$
inside a square. The $M$ mobiles within it are located according to a uniform
clustering distribution with each mobile surrounded by an \emph{exclusion zone} \cite{torval2} with
radius $r_{ex}.$

\subsection{Propagation Model}

At millimeter-wave frequencies, blockages by various obstacles often prevent LOS propagation, but reflections allow a non-line-of-sight (NLOS) propagation of multipath clusters to reach the receiver. {Extensive measurement campaigns, such as those in \cite{rapp}, classify measurements into LOS and NLOS categories, and find separate path-loss exponents and shadowing standard deviations for each category.}   {Building upon these measurements, stochastic millimeter-wave propagation models (e.g., \cite{dire,bai,tal,venu}) characterize the blockages with a stochastic model, and then use random shape theory to determine the probability of LOS propagation.  Such models show the strong dependence of the LOS probability upon the link length.  In \cite{bai,tal,venu}, a fairly accurate approximation is found by classifying all links shorter than a critical distance (the radius of an \emph{LOS ball}) to be LOS, and all links longer to be NLOS.  Such an approximation hardens the stochastic model into a deterministic one with a sharp transition at the critical distance.  In this paper, we propose a refined deterministic model that allows for a more gradual transition from LOS to NLOS regions, thereby capturing the partial or occasional blocking that may occur at intermediate distances.  The key to the model is the use of }
a nonlinear \emph{tanh} function to capture the statistical dynamics of LOS and NLOS paths in a single function with a {rapid, but not instantaneous,} transition between predominately LOS and NLOS zones.  The rate of the transition can be tuned to match the data {for a given blockage model. For such a model,} the blockage and LOS probabilities do not have to be assumed, links do not have to be identified as either LOS or NLOS, and both shadowing and fading can be seamlessly included.   {The fact that some short paths could have a higher attenuation than longer ones is taken into account by the shadowing, which induces a variability in the received power due primarily to partial blockages.}

The propagation model comprises models for the area-mean power, the local-mean power, and the fading \cite{torr}. The area-mean power, which is the average
received power in the network region, is a function of the distance $d$
between a source and destination. The area-mean path-loss function is
expressed as the attenuation power law
\begin{equation}
f\left(  d\right)  =\left(  \frac{d}{d_{0}}\right)  ^{-\alpha\left(  d\right)
},\text{ \ }d\geq d_{0} \label{eq:fd}%
\end{equation}
where $\alpha\left(  d\right)  $ is the attenuation power-law exponent, and
$d_{0}$ is a reference distance that is {less than or equal to $r_{ex}$}. At
microwave frequencies, the power-law exponent is approximately constant, but
at millimeter-wave frequencies, its distance-dependence must be modeled.

Since the area-mean power is an average power over the network region, the distance-dependent model of the power-law exponent for millimeter-wave frequencies is an average over the LOS and NLOS network links. This model reflects the empirical fact that $\alpha(d)$ differs substantially for LOS and NLOS links, tending toward $\alpha_{min}$ for the usually shorter LOS links and tending toward a much larger $\alpha_{max}$ for the usually longer NLOS links \cite{rapp,rang,akd}. Empirical data, such as that shown in Fig. 2 of \cite{rapp}, indicates that there is a small range of link lengths for which there are significant numbers of both LOS and NLOS links within the network region.
Therefore, $\alpha(d)$ is modeled as a monotonically increasing function:
\begin{equation}
\alpha\left(  d\right)  =\alpha_{\min}+\left(  \alpha_{\max}-\alpha_{\min
}\right)  \tanh\left(  \mu d\right)  \label{eq:alpha}%
\end{equation}
which indicates that $\alpha_{\min}\leq\alpha\left(  d\right)  <\alpha_{\max
}.$ The \emph{transition parameter} $\mu$, which can be determined by fitting
to empirical data, controls the transition rate from $\alpha_{\min}$ to a
value close to $\alpha_{\max}$.  \

The local-mean received power at one end of a communication link is the
product of the area-mean power and a factor due to large-scale terrain effects
in the vicinity of the link. This factor is expressed as $10^{\xi/10},$ where
$\xi$ is the \textit{shadowing factor }for the link. The shadowing factor can
be derived from a deterministic terrain model or can be modeled as a random
variable in a statistical model. In this paper, we assume lognormal shadowing
in which the shadowing factors are independent, identically distributed,
zero-mean Gaussian random variables with a distance-dependent variance.

For millimeter-wave frequencies, empirical data \cite{rang,akd,rapp} indicates that the standard deviation of the shadowing factor differs substantially for LOS and NLOS links, tending toward $\sigma_{\min}$ for the usually shorter LOS links and tending toward a much larger $\sigma_{\max}$ for the usually longer NLOS links \cite{rang,akd,rapp}.
Since there is a small range of link lengths for which there are significant
numbers of both LOS and NLOS links, the standard deviation of the shadowing
factor for millimeter-wave frequencies is modeled as a monotonically
increasing function:
\begin{equation}
\sigma_{s}\left(  d\right)  =\sigma_{\min}+(\sigma_{\max}-\sigma_{\min}%
)\tanh\left(  \mu d\right)
\end{equation}
which indicates that $\sigma_{\min}\leq\sigma_{s}\left(  d\right)
<\sigma_{\max}.$

The fading is assumed to have a Nakagami distribution function.
Since the received power for the longer NLOS links is due to multipath clusters, the fading becomes more severe. Therefore, in analogy with (\ref{eq:fd}) and (\ref{eq:alpha}), a distance-dependent Nakagami parameter is modeled as a monotonically
decreasing function:
\begin{equation}
m\left(  d\right)  =m_{\max}-(m_{\max}-m_{\min})\tanh\left(  \mu d\right)
\label{fade}%
\end{equation}
which indicates that $m_{\min}\leq m\left(  d\right)  <m_{\max}.$ If $m\left(
d\right)  $ is constrained to be an integer, then we set it equal to the
integer closest to the value of the right-hand side of (\ref{fade}).

{This propagation model facilitates simple simulations of
the performance of frequency-hopping millimeter-wave uplinks. The model has been validated for the outdoor urban propagation environment and some frequencies. Validation and the extraction of the model parameter values can be done for other propagation environments and frequencies as measurements become available. The propagation model could be replaced by or combined with more specific propagation models. Although the simulation complexity, and the time required for simulations might increase, the basic methodology described below would remain intact.}

\subsection{Beamforming}

{Because of the high propagation losses, high handover rates, intermittent
connectivity, and power limitations of mobile transmitters at millimeter-wave
frequencies, the antenna arrays at the BSs and mobiles will need to form
highly directional beams for uplink transmissions \cite{shokri,zheng,sun}.}
The beams are fixed in space and each mobile aligns its antenna with one of beams of the base station.
{While spatial multiplexing is an option in 4G networks, the multiple-input
multiple-output (MIMO) dimensions in 5G uplinks should be used for
beamforming, which serves as a type of open-loop multiuser MIMO. Providing
multiple layers to individual mobiles is impractical primarily because of the
difficultly in obtaining and feeding back channel-state information when the
handover rates are high, and the need for digital processing of multiple antenna outputs.}

Densification, high mobility, and the severe impact of blockages at
millimeter-wave frequencies cause frequent handovers and hence the need for
rapid mutual beam alignments. Scanning over many angles and the contamination
of the pilot signals carrying angle-of-arrival information are impediments to
adaptive beams. \emph{Sectorization}, which is the division of BS coverage
into $\zeta>1$ fixed angular sector beams centered at the BS, is used to
reduce beam-alignment delays. Sectorization facilitates power control and the
allocations of frequency-hopping patterns within each sector. Thus, we assume
BS sectorization and adaptive beamforming for the mobiles. At millimeter-wave
frequencies, the beams can be implemented using many antenna elements, perhaps
hundreds, and hence they have narrow beamwidths and very small sidelobes and backlobes.

The scalar $S_{l},$ $l=1,2,...,\zeta C,$ represents the $l$th sector or its
receiver, and the scalar $X_{i},$ $i=1,2,...,M,$ represents the $i$th mobile.
The two-component column vector $\mathbf{S}_{l},$ $l=1,2,...,\zeta C,$
represents the location of the $l$th sector receiver, and the two-component
column vector $\mathbf{X}_{i},$ $i=1,2,...,M,$ represents the location of the
$i$th mobile.

Let ${\mathcal{X}}_{l}\ $denote the set of mobiles served by sector $S_{l}$.
Let $X_{r}\in\mathcal{X}_{j}$ denote a reference mobile that transmits a
desired signal to a reference receiver $S_{j}$. Let $\mathsf{g}(i)$ denote a
function that returns the index of the sector serving $X_{i}$ so that
$X_{i}\in{\mathcal{X}}_{l}$ if $\mathsf{g}\left(  i\right)  =l$. The sector
$S_{\mathsf{g}\left(  i\right)  }$ that serves mobile $X_{i}$ is assumed to be
the one with minimum local-mean path loss when the mainlobe of the transmit
beam of $X_{i}$ is aligned with the sector beam of $S_{\mathsf{g}\left(
i\right)  }.$ Thus, the serving sector has index
\begin{equation}
\mathsf{g}\left(  i\right)  =\mathrm{\arg\max}_{l}\,\left\{  10^{\xi_{i,l}%
/10}f\left(  ||\mathbf{S}_{l}-\mathbf{X}_{i}||\right)  ,\text{ }X_{i}%
\in\mathcal{A}_{l}\right\}\label{assoc}
\end{equation}
where $\xi_{i,l}$ is a \textit{shadowing factor }for the link from $X_{i}$ to
$S_{l}$, $f\left(  \cdot\right)  $ is the area-mean path-loss function, and
$\mathcal{A}_{l}$ denote the set of mobiles \emph{covered} by the sector beam
of $S_{l}$. {The implication of (\ref{assoc}) is that the mobile device will associate its uplink with the sector that receives the strongest signal from the device.  While cell-range expansion in a HetNet might cause the downlink to follow a different association rule, this rule is reasonable for either a single-tier network or a HetNet with a decoupled uplink \cite{boc}.} In the absence of shadowing, the serving sector will be the
receiver that is closest to $X_{i}$. In the presence of shadowing, a mobile device
may actually be associated with a sector that is more distant than the closest
one if the shadowing conditions are sufficiently better.

Each sector-beam gain pattern is modeled by a two-level function with maximum
constant gain over a mainlobe with beamwidth equal to $2\pi/\zeta$, and
minimum constant gain over the sidelobes and backlobes. Let $A_{s}$ denote the
average gain of a sector beam. Then the \textit{sector-beam gain pattern}
associated with $S_{l}$ may be expressed as
\begin{equation}
B_{l}\left(  \theta\right)  =\left\{
\begin{array}
[c]{cc}%
A_{s}[b+\zeta\left(  1-b\right)  ], & \psi_{l}\leq\theta\leq\psi_{l}%
+2\pi/\zeta\\
A_{s}b, & \text{otherwise}%
\end{array}
\right.
\end{equation}
$\psi_{l}$ is the offset angle of the beam pattern, and $b$ is the relative
sidelobe and backlobe level.

\begin{figure}[ptb]
\centering
\includegraphics[width=7.5 cm]{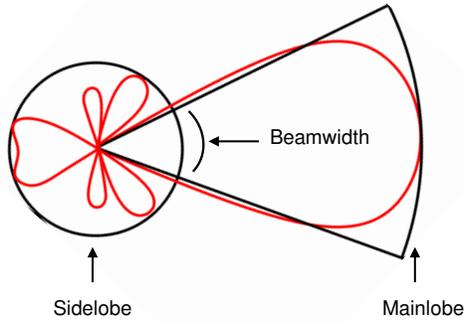} \caption{
{Antenna model (See also Fig1.b of \cite{bai})}. }%
\label{Antenna Model}%
\end{figure}

\begin{table}[ptb]
\begin{tabular}
[c]{|c|c|c|}\hline
& Mobile & BS sector\\\hline\hline
Beamwidth & $\Theta$ & $\displaystyle \frac{2 \pi}{\zeta}$\\\hline\hline
Mainlobe gain & $A_{m}[a+\frac{2\pi\left(  1-a\right)  }{\Theta}]$ &
$A_{s}[b+\zeta\left(  1-b\right)  ]$\\\hline\hline
Sidelobe gain & $A_{m}a$ & $A_{s}b$\\\hline
\end{tabular}
\caption{{Summary of the antenna characteristics.}}%
\label{Summary_antenna}%
\end{table}

Each active mobile points its adaptive beam toward its associated BS.
{The antenna model, which is used to approximate the beamforming patterns, is illustrated in Fig. \ref{Antenna Model}, while the values of beamwidth, mainlobe gain, and sidelobe gain are provided for both the mobile and base station in Table \ref{Summary_antenna}.}
As in \cite{bai}, the antenna pattern is modeled with two gains: one for the
mainlobe with beamwidth $\Theta$, and another for the sidelobes and backlobes.
Let $A_{m}$ denote the average gain of an adaptive mobile beam. Then the
\emph{mobile-beam gain} \emph{pattern} in the direction of the reference
receiver $S_{j}$ due to the angular offset of this beam pattern may be
expressed as
\begin{equation}
b_{i,j}=\left\{
\begin{array}
[c]{cc}%
A_{m}[a+\frac{2\pi\left(  1-a\right)  }{\Theta}], & \frac{\left(
\mathbf{S}_{j}-\mathbf{X}_{i}\right)  ^{T}\left(  \mathbf{S}_{\mathsf{g}%
\left(  i\right)  }-\mathbf{X}_{i}\right)  }{||\mathbf{S}_{j}-\mathbf{X}%
_{i}||||\mathbf{S}_{\mathsf{g}\left(  i\right)  }-\mathbf{X}_{i}||}%
>\cos\left(  \frac{\Theta}{2}\right)  \\
A_{m}a, & \text{otherwise}%
\end{array}
\right.  \label{gainpattern}%
\end{equation}
where $||\mathbf{\cdot}||$\ is the Euclidean norms, the superscript $T$
denotes the transpose, and $a$ is the relative sidelobe and backlobe level. {The main idea behind (\ref{gainpattern}) is to set the mobile's antenna gain to its mainlobe level when it is facing the reference base station and to set it to its sidelobe level when it is facing away.}


\subsection{Frequency Hopping and Intersector Interference}

Frequency hopping \cite{torr} may be used in SC-FDMA uplink systems to provide
the diversity that will mitigate the effects of frequency-selective fading and
intersector interference. Because of network synchronization and similar
propagation delays for the mobiles associated with a cell sector, synchronous
orthogonal frequency-hopping patterns can be allocated so that at any given
instant in time, there is no \emph{intrasector} interference. The
frequency-hopping patterns transmitted by mobiles in other sectors are not
generally orthogonal to the patterns in a reference sector, and hence produce
\emph{intersector} interference. The varying propagation delays from the
interfering mobiles cause their frequency-hopping signals to be asynchronous
with respect to the desired signal. Duplexing prevents uplink interference
from downlink transmissions.

Each mobile uses a frequency-hopping pattern over a hopset with $L$ disjoint
frequency channels. Let $L_{l},$ $l=1,2,\ldots,\zeta C,$ denote positive
integer divisors of $L$ such that $L/L_{l}\geq2.$ Each mobile in
${\mathcal{X}}_{l}\ $is assigned a distinct block of $L_{l}$ contiguous
frequency channels during each of its hop intervals, and the block may change
to any of $L/L_{l}$ disjoint spectral regions with every hop. Consider an
uplink \emph{reference signal} that traverses a reference link from a
reference mobile $X_{r}$ to a reference receiver $S_{j}.$ To minimize edge
effects, the reference mobile is assumed to be located in the yellow zone in
the center of Fig. \ref{fig:Figure_1}. Because of a possible incomplete
spectral overlap, the received interference power from mobile $X_{i}$ at
$S_{j}$ when the mobile's signal collides with the reference signal is reduced
by the \textit{spectral factor}
\begin{equation}
F_{l}=\min\left(  L_{j}/L_{l},1\right)  .
\end{equation}

\begin{figure}[t]
\centering
\includegraphics[width=13.5 cm]{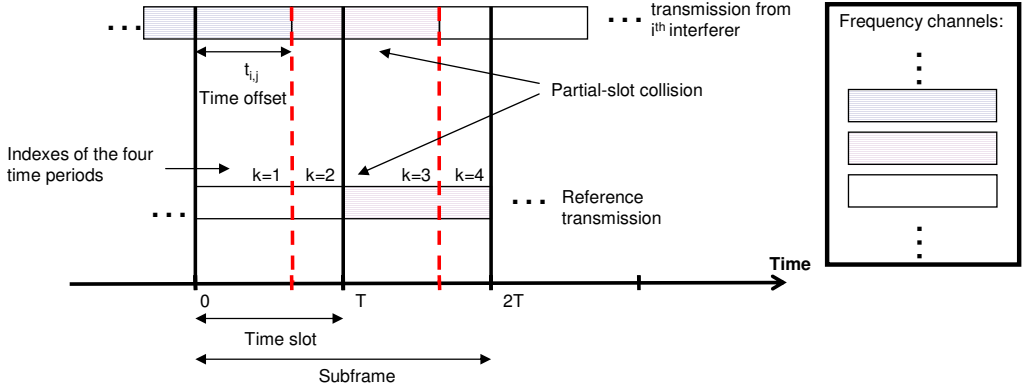}
\caption{Illustration of co-channel interference at sector receiver $S_{j}$ produced by an interfering mobile $X_{i}$ and arriving with a
relative timing offset of $t_{i,j}$. Each block in the illustration represents
a hop, and the selected channel is indicated by the shading of the block. A
partial-slot collision is indicated, where the interfering mobile has selected
the frequency channel used by the reference mobile during the second slot of
the subframe. However, due to the asynchronism, the collision only extends for
part of the slot.}%
\label{fig:Figure_2}%
\end{figure}

Associated with each potentially interfering mobile, which is assumed to
transmit throughout a subframe or not at all, is a hop transition time
$t_{i,j}$ at $S_{j}$ relative to the hop transition time of a pair of hop
intervals of the reference signal. Fig. \ref{fig:Figure_2} illustrates the
relative timing of the signals from the reference mobile $X_{r}$ and the
interfering mobile $X_{i}.$ We assume a static topology, no handoffs,
continual transmissions, and fixed frequency assignments throughout the
duration of the subframe. The reference mobile transmits a turbo codeword of
duration $2T,$ which is aligned with the subframe. It is assumed that the
frequency separation of the two frequency channels of the two slots is
sufficient for independent fading of fixed amplitude in each slot. If the
sector receivers and mobiles are synchronized, then%
\begin{equation}
t_{i,j}=[(||\mathbf{S}_{j}-\mathbf{X}_{r}||-||\mathbf{S}_{j}-\mathbf{X}%
_{i}||)/c]\operatorname{mod}T
\end{equation}
where $c$ is the speed of an electromagnetic wave. As illustrated in Fig.
\ref{fig:Figure_2}, the reference signal encounters four time periods of
potential interference from an active mobile $X_{i}$: $0\leq t\leq t_{i,j},$
$t_{i,j}\leq t\leq T,$ $T\leq t\leq t_{i,j}+T,$ and $t_{i,j}+T\leq t\leq2T$.
The generic index $k\in\{1,2,3,4\}$, denotes a time period of the subframe
with duration that varies with each $t_{i,j}.$ In the example provided in Fig.
\ref{fig:Figure_2}, co-channel interference occurs during the third time
period, where the interfering mobile has selected the frequency channels used
by the reference mobile. The \textit{fractional duration} of each of the four
subframe time periods relative to the subframe period $2T$ are
\begin{equation}
C_{i,j,k}=\left\{
\begin{array}
[c]{cc}%
\frac{t_{i,j}}{2T}, & k=1,3\\
\frac{T-t_{i,j}}{2T}, & k=2,4
\end{array}
\right.  .
\end{equation}

The set of indices of potentially interfering mobiles is $\mathcal{S}=\left\{
i:X_{i}\notin{\mathcal{X}}_{j}\right\}  $. Let $N_{l}$ denote the number of
mobiles associated with sector $S_{l}.$ Because of the required orthogonality
of frequency blocks assigned to mobiles within each sector, $N_{l}\leq$
$L/L_{l}$ and any additional mobiles within the sector are reassigned to other
sectors.
{Each additional mobile is served by the sector from which it receives the next highest power. If all the sectors happened to have excess mobiles, then some mobiles would be denied service.}
In view of the potential spectral overlaps, the maximum number of interfering
mobiles within a sector during a subframe time period is $\min[\max\left(
L_{j}/L_{l},1\right)  ,N_{l}].$ Let $\mathcal{S}_{k}\subset$ $\mathcal{S}$
denote the set of interfering mobiles during subframe time period $k.$ If
$N_{l}\leq\max\left(  L_{j}/L_{l},1\right)  ,$ then all $N_{l}$ mobiles in
sector $l$ are in $\mathcal{S}_{k}$. If $N_{l}>\max\left(  L_{j}%
/L_{l},1\right)  ,$ then some of the mobiles in sector $l$ cannot cause
interference during subframe time period $k$. In that case, we approximate by
randomly selecting a subset of the $N_{l}$ mobiles to be included in
$\mathcal{S}_{k}.$

Let $q_{i,k}$ denote the probability that the signal from a potentially
interfering mobile collides with the reference signal during subframe time
period $k$, $1\leq k\leq4.$ The \emph{activity\ probability} $p_{i}$ is the
probability that mobile $X_{i}$ transmits throughout the time interval
$[0,2T)$. Assuming uniformly distributed frequency-hopping patterns that are
orthogonal within each sector,
\begin{equation}
q_{i,k}=\frac{\max\left(  N_{\mathsf{g}\left(  i\right)  }L_{\mathsf{g}\left(
i\right)  },L_{j}\right)  }{L}p_{i},\text{ \ }i\in\mathcal{S}_{k},\text{
\ }1\leq k\leq4
\end{equation}
and $q_{i,k}=0,$ otherwise$.$\

\subsection{SINR}

The instantaneous signal-to-interference-and-noise ratio (SINR) at sector
receiver $S_{j}$ when the desired signal is from $X_{r}\in\mathcal{X}_{j}$
fluctuates because potentially interfering signals do not always coincide with
the reference signal in time or frequency. Pilot sequences are used to
estimate the complex fading amplitudes in the receiver. Therefore, the
performance of the reference receiver is primarily a function of the
\emph{average SINR} defined as the ratio of the average power of the signal to
the average power of the noise and interference, where the average is over the
time interval of a subframe and turbo codeword. It is assumed that the
adaptive beam of reference mobile $X_{r}\ $is perfectly aligned with the
sector beam. The average SINR during a subframe is%
\begin{equation}
\gamma_{r,j}=\frac{\overline{\rho}_{r,j}}{{\mathcal{N}}+\sum_{k=1}^{4}%
\sum_{i\in\mathcal{S}_{k}}I_{i,k}\rho_{i,j,k}C_{i,j,k}} \label{SINR1}%
\end{equation}
where $\mathcal{N}$ is the noise power, $\overline{\rho}_{r,j}$ is the average
received power from reference mobile $X_{r}$, and $\rho_{i,j,k}$ is the
received power from an interference signal that collides with the reference
signal during subframe time period $k$. The indicators $I_{i,k}$ are Bernoulli
random variables with probabilities%
\begin{align}
P[I_{i,k}  &  =1]=q_{i,k},\text{ }P[I_{i,k}=0]=1-q_{i,k}\nonumber\\
\text{\ }i  &  \in\mathcal{S}_{k},\text{ \ }1\leq k\leq4.
\end{align}

Let $g_{r,j,1}$ and $g_{r,j,2}$ denote the unit-mean power gains due to the
independent fading of the frequency-hopping reference signal in subframe slots
1 and 2, respectively. The power gain of independent Nakagami fading with
parameter $m_{0}=m_{r,j}\ $in each slot has the gamma density function:
\begin{equation}
h_{r,j}(x)=\frac{m_{0}^{m_{0}}x^{m_{0}-1}\exp\left(  -m_{0}x\right)  }%
{\Gamma\left(  m_{0}\right)  }u\left(  x\right)  \label{2}%
\end{equation}
where $u\left(  x\right)  =1,$ $x\geq0,$ and $u\left(  x\right)  =0,$
otherwise. The average power gain due to fading is $\overline{g}_{r,j}=\left(
g_{r,j,1}+g_{r,j,2}\right)  /2.$ Using (\ref{2}), we obtain the density
function of $\overline{g}_{r,j}:$%
\begin{equation}
h_{0}(x)=\frac{\left(  2m_{0}\right)  ^{2m_{0}}x^{2m_{0}-1}\exp\left(
-2m_{0}x\right)  }{\Gamma\left(  2m_{0}\right)  }u\left(  x\right)
\end{equation}
which is the power of a Nakagami random variable with parameter $2m_{0}$. This
doubling of the Nakagami parameter indicates the beneficial effect of the
frequency hopping in mitigating frequency-selective fading. Let ${P}_{r}$ denote the power from $X_{r}$ that would be received at the
reference distance $d_{0}$ with maximum antenna-pair gain in the absence of
shadowing and fading. The average received power from reference mobile $X_{r}$
is%
\begin{equation}
\overline{\rho}_{r,j}={P}_{r}\overline{g}_{r,j}10^{\xi_{r,j}/10}f\left(
d_{r}\right)  \label{pa}%
\end{equation}
where $d_{r}=||\mathbf{S}_{j}-\mathbf{X}_{r}||$ is the length of the reference link.

Let $g_{i,j,k}$ denote the fading gain of the signal from mobile $X_{i}$ at
$S_{j}$ during time interval $k$. Assuming that the bandwidths of the
$L/L_{l}$ and $L/L_{j}$ disjoint spectral regions exceed the coherence
bandwidth, the \{$g_{i,j,k}\}$ are independent for each hop interval with
unit-mean, and $g_{i,j,k}=a_{i,j,k}^{2}$, where $a_{i,j,k}$ has a Nakagami
distribution with distance-dependent parameter $m_{i,j}$. Let ${P}_{i}$ denote
the power from $X_{i}$ that would be received at the reference distance
$d_{0}$ with maximum antenna-pair gain in the absence of shadowing and fading.
Allowing for the spectral and beam factors, the received power from $X_{i}$ at
$S_{j},$ $i\in\mathcal{S}_{k},$ during time interval $k$ is%
\begin{align}
\rho_{i,j,k}  &  ={P}_{i}g_{i,j,k}10^{\xi_{i,j}/10}f\left(  ||\mathbf{S}%
_{j}-\mathbf{X}_{i}||\right)  F_{\mathsf{g}\left(  i\right)  }\frac
{b_{i,j}B_{j}\left(  \theta_{i,j}\right)  }{B_{\max}}\nonumber\\
\text{\ }i  &  \in\mathcal{S}_{k},\text{ \ }1\leq k\leq4 \label{pi}%
\end{align}
where {the \emph{maximum antenna-pair gain} is%
\begin{equation}
B_{\max}=A_{s}A_{m}\left[  b+\zeta\left(  1-b\right)  \right]  \left[
a+\frac{2\pi\left(  1-a\right)  }{\Theta}\right]
\end{equation}
and $\theta_{i,j}$ is the arrival angle at $S_{j}$ of a signal from $X_{i}$.}

\subsection{Power Control}

The synchronous, orthogonal uplink signals in a sector ensure the absence of
near-far problems and intrasector interference in principle. However, power
control is needed to limit the potential intrasector interference caused by
synchronization errors and hardware imperfections. Closed loop uplink power
control may be implemented by monitoring the received powers of the SC-FDMA
signals at the sector receiver, and then feeding back control commands to the
sector mobiles that constrain the local-mean power received from each sector
mobile, {which is assumed
to have its beam aligned with its associated sector beam. To provide mobiles with some flexibility in exploiting favorable channel conditions, {\em fractional power control} of the constrained local-mean power implies that
\begin{eqnarray}
P_{r}\left[10^{\xi_{r,j}/10}f\left( d_{r}\right)
\right]^{\delta}\hspace{-0.25cm}&=&\hspace{-0.25cm}P_{i}\left[ 10^{\xi_{i,\mathsf{g}\left(  i\right)  }/10}f\left(  ||\mathbf{S}_{\mathsf{g}\left(  i\right)
}-\mathbf{X}_{i}||\right)\right]^{\delta},  \nonumber \\
& & \text{ \ }i  \in\mathcal{S}_{k}, 0 < \delta < 1 \label{pc}
\end{eqnarray}
where $\delta$ is the {\em power-control parameter}. If $\delta= 0$, there is no power control, and transmitter powers are all equal. If $\delta =1$, full power control forces the received local-mean powers from all
mobiles to be equal. If $0 < \delta < 1$, then decreasing $\delta$ improves the performance of some mobiles in a sector while increasing the interference in neighboring sectors.}

Substituting (\ref{pa}), (\ref{pi}), and (\ref{pc}) into (\ref{SINR1}), we
obtain
\begin{equation}
\gamma_{r,j}=\frac{\overline{g}_{r,j}}{\Gamma_{0}^{-1}+\sum_{k=1}^{4}\sum_{i\in\mathcal{S}_{k}}I_{i,k}\Omega_{i,j}g_{i,j,k}C_{i,j,k}}
\end{equation}
where
\begin{equation}
\Omega_{i,j}=\frac{10^{[\xi_{i,j}-\delta\xi_{i,\mathsf{g}\left(  i\right)}+\left( \delta-1\right)\xi_{r,j}
]/10}f\left(  ||\mathbf{S}_{j}-\mathbf{X}_{i}||\right) F_{\mathsf{g}\left(
i\right)  }b_{i,j}B_{j}\left(  \theta_{i,j}\right)}{
\left[f \left( d_{r}\right)\right]^{1-\delta}
\left[f\left(  ||\mathbf{S}
_{\mathsf{g}\left(  i\right)  }-\mathbf{X}_{i}||\right)\right]^{\delta} B_{\max}}
\label{omega}
\end{equation}
 is the ratio of the interference power from
$X_{i}$ to the reference-signal power, and%
\begin{equation}
\Gamma_{0}=\frac{P_{r}}{\mathcal{N}}10^{\xi_{r,j}/10}f\left(  d_{r}\right)
\label{gam}%
\end{equation}
is the signal-to-noise ratio (SNR) at the sector receiver in the absence of
fading. Because of the division in (\ref{omega}), $\Omega_{i,j}$ does not
depend on the average gains $A_{s}$ and $A_{m}.$ {The \emph{reference SNR} is
$P_{r}/\mathcal{N},$ which is the received SNR at distance $d_{0}$ when the
maximum antenna-pair gain occurs and shadowing and fading are absent.}

\section{Outage Probability}

\label{sec:Outage_Probability}

Let $\beta$ denote the minimum average SINR required for reliable reception of
a signal from $X_{r}$ at its serving sector receiver $S_{j},$ $j=\mathbf{g}%
(r)$. An \emph{outage} occurs when the average SINR of a signal from $X_{r}$
falls below $\beta$. The value of $\beta$ sets a limit on the code-rate $R$ of
the uplink, which is expressed in units of bits per channel use (bpcu), and
depends on the modulation and coding schemes, and the overhead losses due to
pilots, cyclic prefixes, and equalization methods. The exact dependence of $R$
on $\beta$ can be determined empirically through tests or simulation.

The set of $\{\Omega_{i,j}\},$ $i\in\mathcal{S}_{k},$ for reference receiver
$S_{j}$ is represented by the vector $\boldsymbol{\Omega}_{j}$. Conditioning
on $\boldsymbol{\Omega}_{j}$, the \emph{outage probability} of a desired
signal from $X_{r}\in{\mathcal{X}}_{j}$ that arrives at $S_{j}$ is
\begin{equation}
\epsilon=P\left[  \gamma_{r,j}\leq\beta\big|\boldsymbol{\Omega}_{j}\right]  .
\end{equation}
Because it is conditioned on $\boldsymbol{\Omega}_{j}$, the outage probability
depends on the particular network realization, which has dynamics over
timescales that are much slower than the fading or frequency hopping. We
define
\begin{equation}
\beta_{0}=2\beta m_{0},\text{ \ }z=\Gamma_{0}^{-1}%
\end{equation}
where the Nakagami parameter $m_{0}=\left[  m_{r,j}\right]  $ for the
reference uplink signal is assumed to be a positive integer. A derivation
similar to the one in \cite{torval1} yields
\begin{equation}
\epsilon=1-e^{-\beta_{0}z}\sum_{s=0}^{2m_{0}-1}{\left(  \beta_{0}z\right)
}^{s}\sum_{t=0}^{s}\frac{z^{-t}}{(s-t)!}H_{t}(\boldsymbol{\Omega})\text{ }
\label{a1}%
\end{equation}
where%
\begin{equation}
H_{t}(\boldsymbol{\Omega})={\displaystyle\sum\limits_{\substack{\ell_{ik}%
\geq0\\\sum_{k=1}^{4}\sum_{i\in\mathcal{S}_{k}}\ell_{ik}=t}}} \prod_{k=1}%
^{4}\prod_{i\in\mathcal{S}_{k}}G_{\ell_{ik}}(i,j,k) \label{a2}%
\end{equation}
the summations in (\ref{a2}) are over all sets of nonnegative indices that sum
to $t$,%
\begin{equation}
G_{\ell}(i,j,k)=%
\begin{cases}
1-q_{i,k}(1-\Psi_{i,j,k}^{m_{i,j}}), & \ell=0\\
\frac{q_{i,k}\Gamma(\ell+m_{i,j})}{\ell!\Gamma(m_{i,j})}\left(  \frac
{\Omega_{i,j}C_{i,j,k}}{m_{i,j}}\right)  ^{\ell}\Psi_{i,j,k}^{m_{i,j}+\ell} &
\ell>0
\end{cases}
\end{equation}
and%
\begin{equation}
\Psi_{i,j,k}=\left(  \beta_{0}\frac{\Omega_{i,j}C_{i,j,k}}{m_{i,j}}+1\right)
^{-1}\hspace{-0.5cm},\text{ \ }i\in\mathcal{S}_{k},\text{ \ }1\leq k\leq4.
\label{a4}%
\end{equation}

\section{Numerical Results}

\label{sec:Results}

In the following examples, {statistical} performance metrics are calculated by
using a Monte Carlo approach with $N$ simulation trials. In each simulation
trial, a realization of the network of Fig. 1 is obtained by placing $M$
mobiles within it according to a uniform clustering distribution with
$r_{ex}=d_{0}=0.004$ km. Randomly generated shadowing factors are used in the
association of mobiles with cell sectors and in other computations. Independent shadowing is assumed among all uplinks, an assumption that becomes increasingly accurate as densification limits the number of mobiles with beams pointing toward a sector beam.

The code rate permitted by the threshold is given by%
\begin{equation}
R=\log_{2}\left(  1+l_{s}\beta\right)
\end{equation}
where $l_{s}=0.794$ corresponds to a 1 dB loss relative to the Shannon bound
for complex discrete-time AWGN channels. The density of mobiles in a network
of area $A_{\mathsf{net}}$ is $\lambda=M/A_{\mathsf{net}}.$ The outage
probability $\epsilon_{i}$ of the reference link for simulation trial $i$ is
computed by applying (\ref{a1})$-($\ref{a4}). The \emph{throughput} of the
reference uplink for simulation trial $i$ is $R(1-\epsilon_{i})$. The average
outage probability over all the simulation trials is
\begin{equation}
\overline{\epsilon}=\frac{1}{N}\sum\limits_{i=1}^{N}\epsilon_{i}.
\end{equation}
The maximum rate of successful data transmissions per unit area is
characterized by the \emph{area spectral efficiency}, defined as
\begin{equation}
\mathcal{A}=\lambda R\left(  1-\overline{\epsilon}\right)  \label{a8}%
\end{equation}
where the units are bits per channel use per unit area.

To avoid edge effects, the performance is measured for a reference mobile
placed in the yellow shaded area of Fig. \ref{fig:Figure_1}, which is a
1 km by 1 km square located in the middle of the network. To
consider the effect of BS densification, the relative topology of the BSs and
the density of mobiles {served within each subframe} $\lambda=100/km^{2}$ are maintained while the size of
the network is scaled (dimensions redefined). For a fixed number of BSs, each
scaling changes the number of mobiles per sector and the transmission
distances. In considering the effects of intersector interference, only the
strongest 30 signals were used, as the attenuation of further signals was
severe enough to make their individual effects negligible.

\begin{figure}[t]
\centering
\includegraphics[width=10.5 cm]{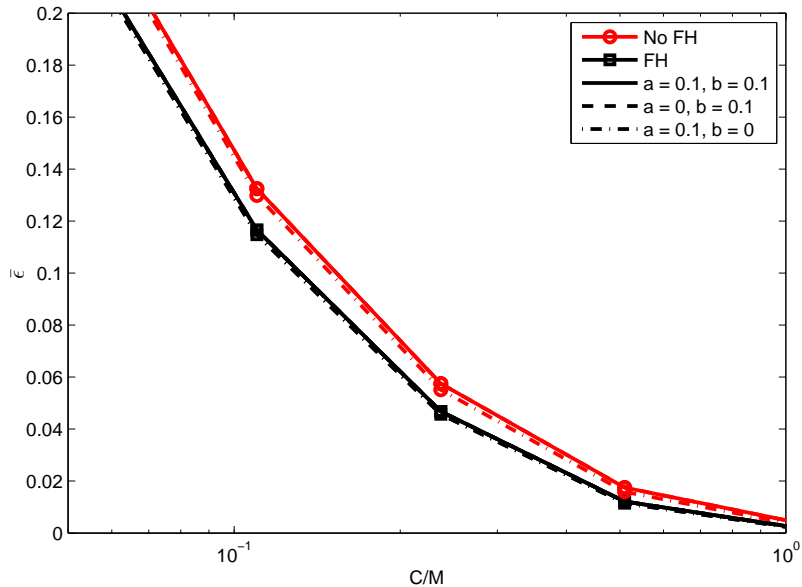} \vspace{-0.25cm}
\vspace{-0.25cm}\caption{
{Average outage probability as a function of densification for both
frequency hopping and its absence.  Densification is quantified by the average number of base stations per mobile.  In this example, a user density of $\lambda = 100$ mobiles per $\mathsf{km}^2$ is assumed.}
}%
\label{fig:Figure_3}%
\end{figure}

\begin{figure}[t]
\centering
\includegraphics[width=10.5 cm]{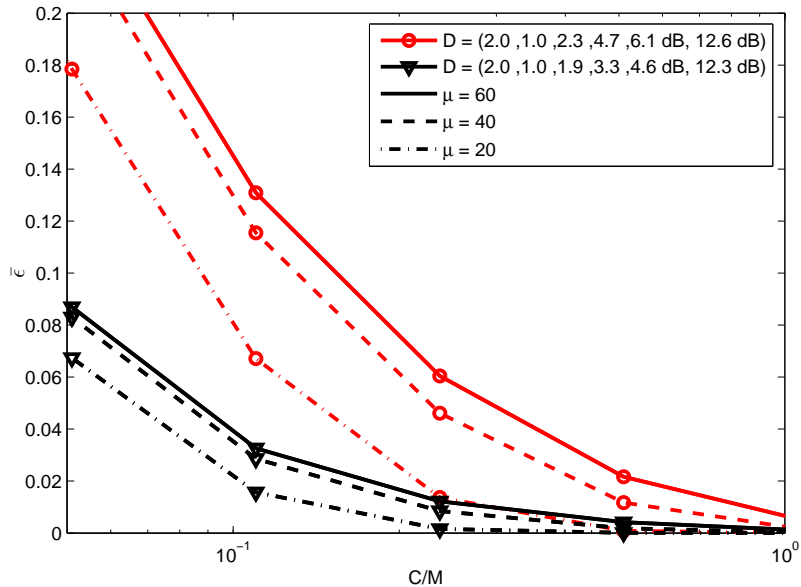} \vspace{-0.25cm}
\vspace{-0.25cm}\caption{
{Average outage probability for several values of the transition
parameter $\mu$ and two sets of propagation parameters. \label{fig:Figure_4}}}%
\end{figure}

\begin{figure}[t]
\centering
\includegraphics[width=10.5 cm]{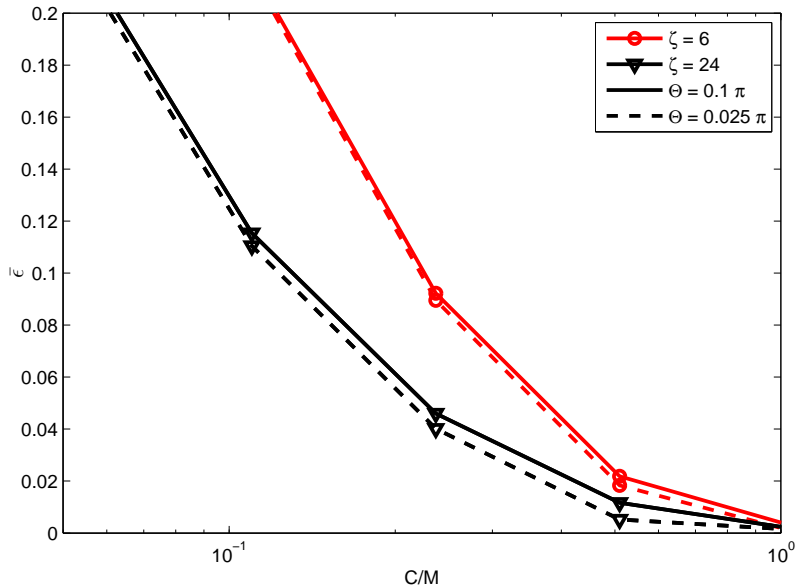} \vspace{-0.25cm}
\vspace{-0.25cm}\caption{
{Average outage probability for several values of the mobile-beam
beamwidth $\Theta$ and the number of sectors $\zeta.$} }%
\label{fig:Figure_5}%
\end{figure}

\begin{figure}[t]
\centering
\includegraphics[width=10.5 cm]{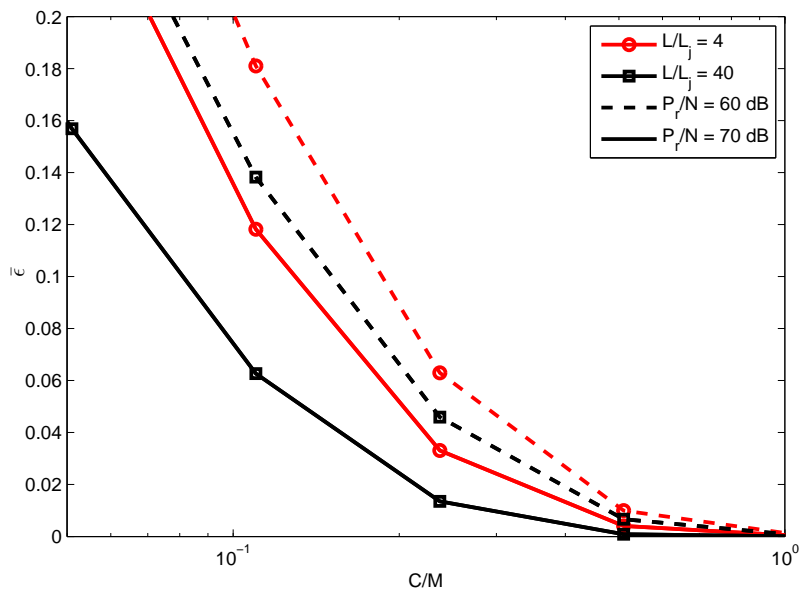} \vspace{-0.5cm}
\caption{
{Average outage probability for several values of $L/L_{j}$ and
$P_{r}/\mathcal{N}$.} }%
\label{fig:Figure_6}%
\end{figure}

\begin{figure}[t]
\centering
\includegraphics[width=10.5 cm]{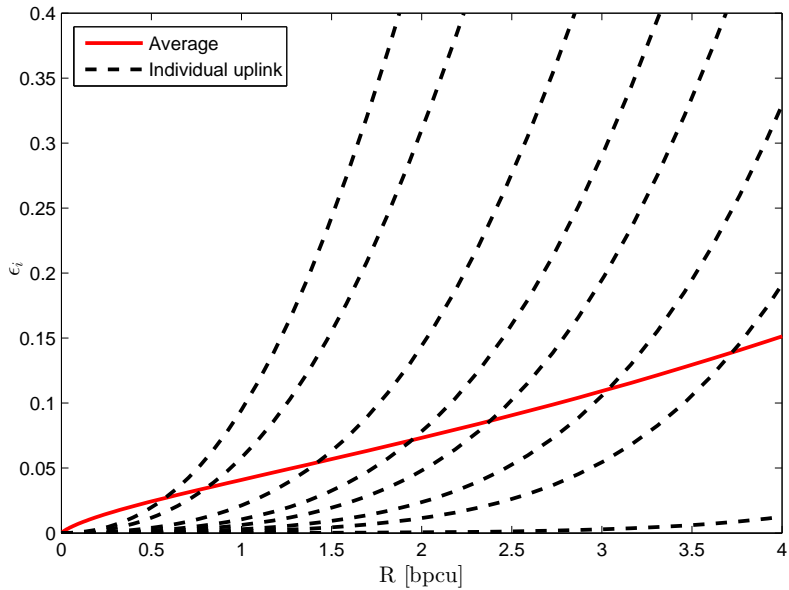} \vspace{-0.25cm}
\vspace{-0.25cm}\caption{
{Outage probabilities of eight uplinks and the average outage
probability over all the uplinks as funtions of the code rate for $C/M=0.1$
and a single simulation trial.}}%
\label{fig:Figure_7}%
\end{figure}

\begin{figure}[t]
\centering
\includegraphics[width=10.5 cm]{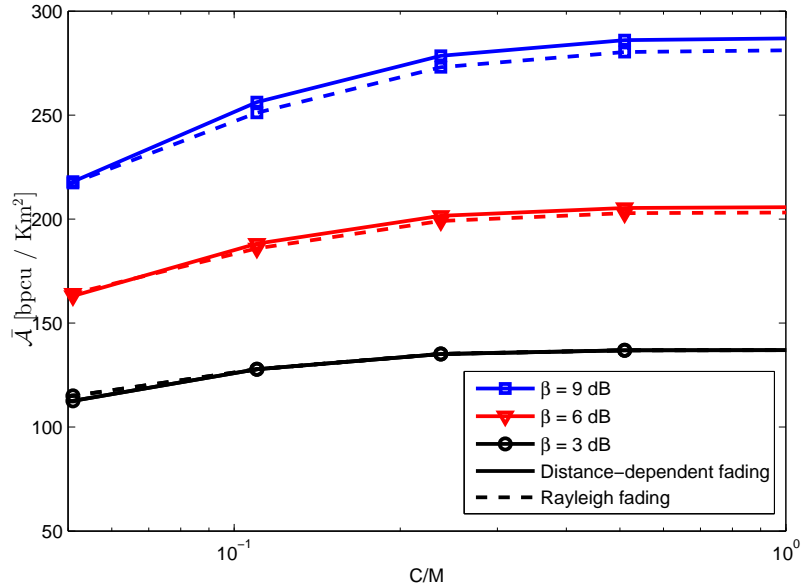} \vspace{-0.25cm}
\vspace{-0.25cm}\caption{
{Area spectral efficiency for three values of the SINR threshold and
both distance-dependent and Rayleigh fading}}%
\label{fig:Figure_8}%
\end{figure}

\begin{figure}[t]
\centering
\includegraphics[width=10.5 cm]{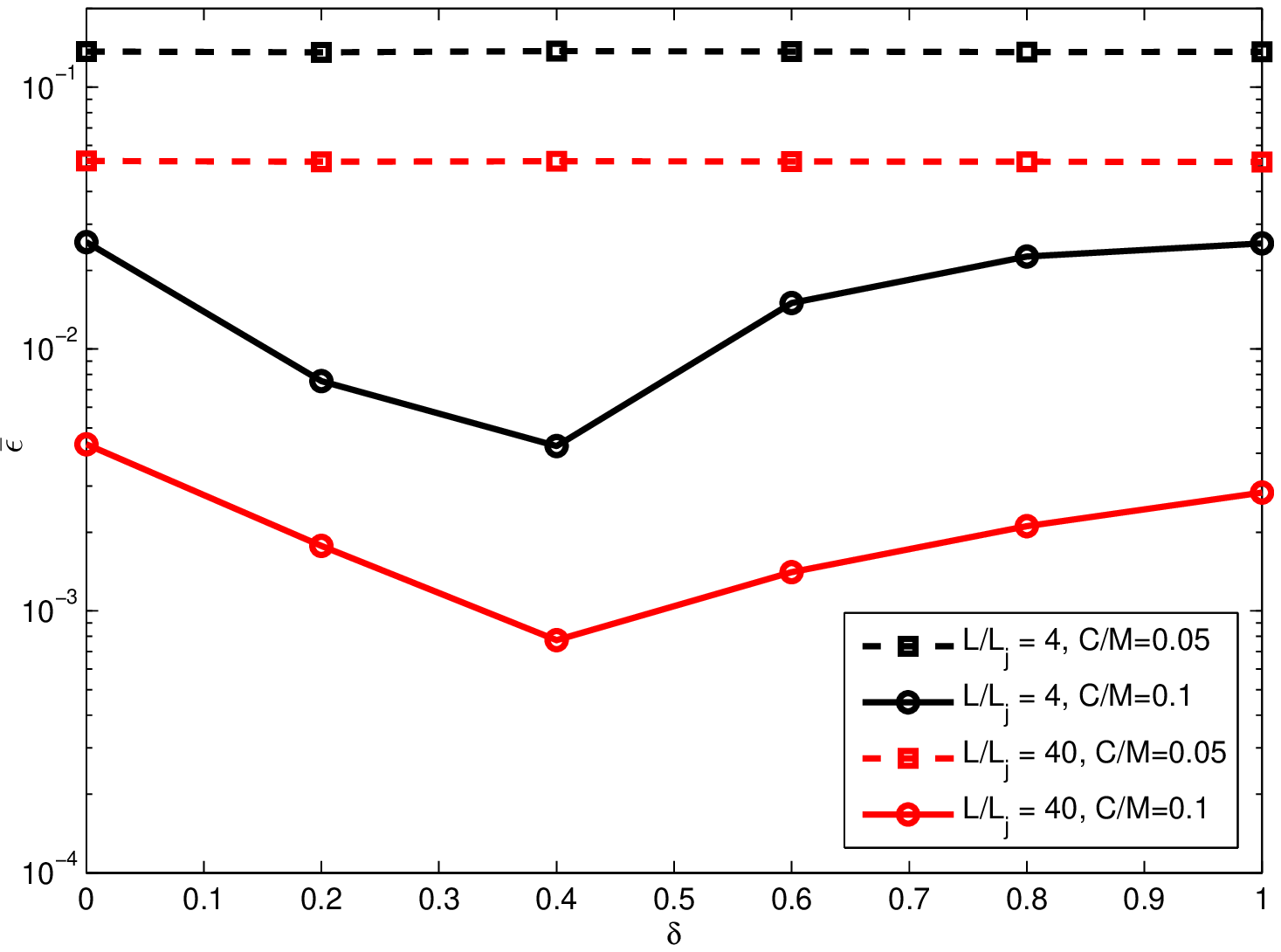} \vspace{-0.25cm}\caption{
{Average outage probability as a function of the power-control parameter.}}%
\label{fig:Figure_9}%
\end{figure}

The ratio $C/M$ serves as a measure of the densification of the BSs. To
capture the impact of densification, we take into account the decrease in the
\emph{typical link-length }$d_{r}$ of the reference link and the consequent
increase in $\Gamma_{0}$ as C/M increases. The typical link-length is defined
as one-fourth of the average separation between BSs. For a fixed density of
mobiles, the average area associated with a cell is inversely proportional to
$C/M$, and hence the average cell radius and a \emph{typical} $d_{r}$ are
proportional to $1/\sqrt{C/M}.$ For the range of C/M of interest, we consider
\begin{equation}
d_{r}=\frac{d_{r0}}{\sqrt{C/M}},\text{\ \ }0.05\leq C/M\leq1 \label{L}%
\end{equation}
where $d_{r0}=25$ m is the typical link length when $C/M=1.$

{Unless otherwise stated the reference SNR in the examples is $P_{r}%
/\mathcal{N}=70$ dB. This SNR might be realized as follows. The transmitter
output power is 30 dBm, the path loss at the reference distance is 90 dB, the
maximum antenna-pair gain is 40 dB, and thus $P_{r}=-20$ dBm. The noise power
spectral density is -174 dBm/Hz, the reference-signal bandwidth is 1 GHz, the
noise figure is 6 dB, and thus $\mathcal{N}=-90$ dBm.}

The slot duration is $T=0.5$ ms, and the common activity factor is $p_{i}=1$.
Unless otherwise stated, frequency hopping is used, and other parameter values
are $L/L_{j}=L/L_{l}=10,$ $\mu=20/km$, $b=0.01,$ $\zeta=24,$ $\Theta=0.1\pi$
radians, $a=0.1$, $\beta=3$ dB, $\delta=0.1$, and $N=10^{5}.$
Unless otherwise stated, the propagation parameters are $D=\left(  m_{\max
},m_{\min},\alpha_{\min},\alpha_{\max},\sigma_{\min},\sigma_{\max}\right)
=\ \left(  2.0,1.0,2.3,4.7,6.1\text{ dB},12.6\text{ dB}\right)  ,$ which are
similar to those expected in a harsh urban environment such as central New
York when transmitting {in the vicinity of 73 GHz \cite{rapp}}.

Figures \ref{fig:Figure_3}-\ref{fig:Figure_6} depict the average outage
probability $\overline{\epsilon}$ for a typical link as a function of the
densification and various parameter values. {The primary reasons for the
monotonic decrease in $\overline{\epsilon}$ with densification are the
increases in $\Gamma_{0}$ and the reduced path loss and shadowing experienced
by the reference signal.} Fig. \ref{fig:Figure_3} illustrates the significant
benefit of frequency hopping. In the absence of frequency hopping, the
reference signal is assumed to experience constant fading over a subframe,
which is valid if subframe duration is less than the coherence time. The
figure shows that the relative decrease in $\overline{\epsilon}$ due to the
use of frequency hopping increases with densification. A viable network would
require an $\overline{\epsilon}$ below 0.05, which in this example requires
$C/M\gtrapprox0.2$ if frequency hopping is used and $C/M\gtrapprox0.35$ if it
is not. The curves for $a=0.1,$ $b=0$ and $a=0,$ $b=0.1$ indicate that
decreasing either the sector-beam sidelobe level below $b=0.1$ or the
mobile-beam sidelobe level below $a=0.1$ has a very small effect on
$\overline{\epsilon}.$

Fig. \ref{fig:Figure_4} compares the effects of several values of the
transition parameter $\mu$ and two sets of propagation parameters. The set
$D=\left(  2.0 ,1.0 ,2.3 ,4.7 ,6.1\text{ dB}, 12.6\text{ dB}\right)  $ is
based on data for New York, whereas the set $D=\left(  2.0, 1.0, 1.9, 3.3,
4.6\text{ dB}, 12.3\text{ dB}\right)  $ is based on data for Austin, Texas
\cite{rapp}. The figure illustrates the benefits of low transition rates and
less congested urban networks. The six curves in the figure converge toward
each other as the densification increases because then the typical link lies
in the LOS region with similar parameter values for both New York and Austin.

Fig. \ref{fig:Figure_5} compares the effects of several values of the mobile
beamwidth $\Theta$ and the number of sectors $\zeta.$ A decrease in $\Theta$
is of minor importance, whereas an increase in $\zeta$ is much more important,
particularly for high levels of densification. {In this figure, $P_{r}%
/\mathcal{N}=70$ dB is maintained for all curves. However, if instead the
average gains of the antennas are maintained, then the maximum antenna-pair
gain increases with the increase in the number of sectors because of their
narrower beamwidths. As a result, $P_{r}/\mathcal{N}$ increases, and there is
a further decrease in the average outage probability.}

Fig. \ref{fig:Figure_6} compares the effects of several values of $L/L_{j}$
and $P_{r}/\mathcal{N}$. {For a fixed $P_{r}/\mathcal{N},$ a decrease in
$L/L_{j}$ corresponds to a decrease in the total available bandwidth relative
to the reference-signal bandwidth. The figure shows that decreasing $L/L_{j}$
by an order of magnitude causes a substantial increase in $\overline{\epsilon
}$ due to the decreased effectiveness of the frequency hopping in enabling the
reference signal to avoid interference. However, the increase in
$\overline{\epsilon}$ is less significant as the densification increases,
which indicates the diminishing importance of the intercell interference. A
further indication of the predominance of the noise in limiting performance is
the substantial increase in $\overline{\epsilon}$ when $\overline{\epsilon
}>0.02$ and $P_{r}/\mathcal{N}$ is reduced by 10 dB. Since beamforming and
sufficient densification provide a noise-limited performance, there is little
or no need for uplink scheduling to eliminate pilot contamination or
interference among mobiles.}

In Fig. \ref{fig:Figure_7}, outage probabilities are plotted as functions of
the code rate $R$ for $C/M=0.1$ and a single simulation trial. The dashed
lines in the figure were generated by selecting eight random uplinks in the
network and computing each outage probability. The average over all the
uplinks is represented by the solid line. Despite the use of {fractional} power
control, there is considerable variability in the dependence of the outage
probability on the code rate due to the irregular network topology, which
results in cell sectors of variable areas and numbers of mobiles.

Increases in the SINR threshold $\beta$ of the network links increase the
outage probability. However, for sufficiently large values of C/M and hence
$\Gamma_{0},$ this effect is minor compared with increased code rate that can
be accommodated. As a result, the area spectral efficiency $\mathcal{A}$
increases significantly, as illustrated in Fig. \ref{fig:Figure_8}. Although
high code rates lead to high values of area spectral efficiency, they also
lead to high outage probabilities for many network links, as illustrated in
Fig. \ref{fig:Figure_7}. Therefore, a compromise solution in choosing the code
rate will be necessary. Fig. \ref{fig:Figure_8} also illustrates that mild
distance-dependent fading modeled in this paper provides an insignificant
improvement relative to Rayleigh fading.

Figure \ref{fig:Figure_9} illustrates the effect of fractional power control on the average outage probability. The choice of the power-control parameter is largely irrelevant if $C/M < 0.1$. If $C/M \geq 0.1$ the optimal value of the power-control parameter is $\delta\approx 0.4$, but the outage probability is low for all choices of $\delta$.

\section{Conclusions}

\label{sec:Conclusions}

This paper derives an analytical model for calculating the outage probability
and area spectral efficiency for frequency-hopping millimeter-wave uplinks that are strong
candidates for use in 5G networks. The model includes the effects of
millimeter-wave propagation, directional beams, frequency hopping, an
arbitrary network topology, and the assignment of frequency blocks to mobiles.
{This paper applies deterministic geometry, which enables the accommodation of actual
base-station topologies, rather than stochastic geometry, which usually relies on Poisson-distributed ones.} In contrast to other propagation models, the propagation model in this paper is based on the direct application of received-power data.
Numerical examples illustrate the effects of various features and parameters.
The BS densification is shown to have critical importance in achieving good
network performance. The significance of the intercell interference is greatly
reduced because of the highly directional sectorization and beamforming. The
usefulness of frequency hopping in compensating for frequency-selective fading
and reducing interference is illustrated. The benefits of increased
sectorization, {low transition rates, increased available bandwidth,} and less
congested urban networks are illustrated. For a sufficient degree of
densification, the area spectral efficiency increases with the code rate
despite the increase in the outage probability. However, a high code rate
leads to high outage probabilities for many network links, and a compromise
solution in choosing the code rate will be necessary. The minor importance of the power-control parameter of the fractional power control is shown.  While the numerical results in this paper correspond to a representative example network, the methodology is general enough to be extended to a wide class of future-generation wireless networks.


\begin{thebibliography}{99}
\bibitem {andr}J. G. Andrews, S. Buzzi, W. Choi, S. V. Hanly, A. Lozano, A. C. K. Soong, and J. C. Zhang, ``What will 5G be?," \textit{IEEE J. Selec. Areas Commun.}, vol. 32, no. 6, pp. 1065-1082, June 2014.

\bibitem {wang}P. Wang, Y. Li, L. Song, and B. Vucetic, ``Multi-gigabit millimeter wave wireless
communications for 5G: From fixed access to cellular networks,"
\textit{IEEE\ Commun. Mag.}, vol. 53, no. 1, pp. 168-178, Jan. 2015.

\bibitem {deb}S. Deb and P. Monogioudis, ``Learning-based uplink interference
management in 4G LTE cellular systems," \textit{IEEE/ACM Trans. Networking},
vol. 23, pp. 398-411, April 2015.

\bibitem {ali}N. Abu-Ali, A.-E. M. Taha, M. Salah, and H. Hassanein, ``Uplink
scheduling in LTE and LTE-Advanced: Tutorial, survey and evaluation
Framework," \textit{IEEE Commun. Surveys Tuts.}, vol. 16, pp. 1239-1265, third
quarter 2014.

\bibitem {ku}G. Ku and J. M. Walsh, ``Resource allocation and link adaptation
in LTE and LTE Advanced: A Tutorial," \textit{IEEE Commun. Surveys Tuts.},
vol. 17, pp. 1605-1633, third quarter 2015.

\bibitem {boc} F. Boccardi, J. G. Andrews, H. Elshaer, M. Dohler, S. Parkvall, P. Popovski, and S. Singh,
``Why to decouple the uplink and downlink in cellular networks and how to do it,''
\textit{IEEE\ Commun. Mag.}, vol. 54, no. 3, pp. 110-117, Mar. 2016.

\bibitem {tsv}D. Torrieri, S. Talarico, and M. C. Valenti, ``Performance analysis of fifth-generation cellular uplink," in \textit{Proc. IEEE Military Communications Conference (MILCOM)}, (Tampa, FL), Oct. 2015.

\bibitem {torval1}D. Torrieri and M. C. Valenti, \textquotedblleft The outage
probability of a finite ad hoc network in Nakagami fading,\textquotedblright%
\ \textit{IEEE Trans. Commun.}, vol. 60, no, 11, pp. 3509-3518, Nov. 2012.

\bibitem {hae}M. Haenggi, \textit{Stochastic Geometry for Wireless Networks}, Cambridge University Press, 2013.Ми

\bibitem {dire}M. Di Renzo, ``Stochastic geometry modeling and analysis of
multi-tier millimeter wave cellular networks," \textit{IEEE Trans. Wireless
Commun.}, vol. 14, no. 9, pp. 5038-5057, Sept. 2015.

\bibitem {val}M.C. Valenti, D. Torrieri, and S. Talarico, \textquotedblleft A
direct approach to computing spatially averaged outage probability,"
\textit{IEEE Commun. Letters}, vol. 18, no. 7, pp. 1103-1106, July 2014.

\bibitem {li}Y. Li, F. Baccelli, H. S. Dhillon, and J. G. Andrews,
``Statistical modeling and probabilistic analysis of cellular networks with
determinantal point processes," \textit{IEEE Trans. Commun.}, vol. 63, no. 9, pp.
3405-3422, Sept. 2015.

\bibitem {torval2}D. Torrieri and M. C. Valenti, \textquotedblleft Exclusion and guard zones in DS-CDMA ad hoc networks,\textquotedblright%
\ \textit{IEEE Trans. Commun.}, vol. 61, no. 6, pp. 2468-2476, June 2013.

\bibitem {rapp}T. S. Rappaport, G. R. MacCartney, M.K. Samimi, and S. Sun, \textquotedblleft Wideband
Millimeter-wave propagation measurements and channel models for future wireless communication system design," \textit{IEEE Trans. Commun.}, vol. 63, no. 9, pp. 3029-3056, Sept. 2015.

\bibitem {bai}T. Bai, and R. W. Heath Jr., \textquotedblleft Coverage and rate analysis for millimeter wave cellular networks," \textit{IEEE Trans. Wireless Commun.}, vol. 14, no. 2, pp. 1100-1114, Feb. 2015.

    \bibitem {tal} S. Talarico and M. C. Valenti, ``Frequency hopping on a 5G millimeter-wave uplink,'' in \emph{Proc. Asilomar Conf. on Signals, Sys., \& Comp.,}, (Pacific Grove, CA), Nov. 2015.

\bibitem {venu}K. Venugopal, M.C. Valenti, and R. W. Heath Jr., \textquotedblleft Device-to-device millimeter wave communications: Interference, coverage, rate, and finite topologies," \textit{IEEE Trans. Wireless Commun.}, to appear.


\bibitem {torr}D. Torrieri, \textit{Principles of Spread-Spectrum
Communication Systems, 3rd ed.}, Springer, 2015.

\bibitem {rang}S. Rangan, T. S. Rappaport, and E. Erkip, \textquotedblleft
Millimeter-wave cellular wireless networks: Potentials and
challenges,\textquotedblright\ \textit{Proc. IEEE}, vol. 102, no. 3, pp. 366-385,
Mar. 2014.

\bibitem {akd}M. R. Akdeniz, Y. Liu, M. K. Samimi, S. Sun, S. Rangan, T. S. Rappaport, and E. Erkip, ``Millimeter wave channel modeling and cellular capacity evaluation," \textit{IEEE J. Selected Areas Commun.}, vol.
32, no. 6, pp. 1164-1179, June 2014.

\bibitem {shokri}H. Shokri-Ghadikolaei, C. Fischione, G. Fodor, P. Popovski, and M. Zorzi, ``Millimeter wave cellular networks: A MAC layer perspective," \textit{IEEE Trans. Commun.}, vol. 63, no. 10, pp. 3437-3458, Oct. 2015.

\bibitem {zheng}K. Zheng, L. Zhao, J. Mei, B. Shao, W. Xiang, and L. Hanzo, ``Survey of large-scale MIMO systems," \textit{IEEE Commun. Surveys Tuts.}, vol. 17, pp. 1738-1760, third quarter 2015.

\bibitem {sun}S. Sun, T. S. Rappaport, R. W. Heath Jr., A. Nix, and S. Rangan, ``MIMO for millimeter-wave wireless communications: Beamforming, spatial multiplexing, or both?," \textit{IEEE\ Commun. Mag.}, vol. 52, no. 12, pp. 110-121, Dec. 2014.

\end{thebibliography}
\end{document}